\documentclass{article}
\usepackage{amsmath}
\usepackage[dvips]{epsfig}
\usepackage{amssymb}
\usepackage{fullpage}

\def\##1{{\underline #1}}
\def\=#1{\underline{\underline{#1}}}

\def\+#1{\underline{\bf #1}}
\def\*#1{\underline{\underline{\bf #1}}}

\def\eps{\epsilon}

\def\muo{\mu_{\scriptscriptstyle 0}}

\def\.{\mbox{ \tiny{$^\bullet$} }}

\def\le{\left(}
\def\ri{\right)}
\def\les{\left[}
\def\ris{\right]}
\def\lec{\left\{}
\def\ric{\right\}}

\def\l#1{\label{#1}}
\def\r#1{(\ref{#1})}

\def\eps{\epsilon}

\def\muo{\mu_0}

\def\.{\mbox{ \tiny{$^\bullet$} }}

\def\le{\left(}
\def\ri{\right)}
\def\les{\left[}
\def\ris{\right]}
\def\lec{\left\{}
\def\ric{\right\}}

\def\l#1{\label{#1}}
\def\r#1{(\ref{#1})}

\begin{document}

\begin{center} {\bf {\Large The Huygens principle for a uniaxial dielectric--magnetic medium with gyrotropic--like
 magnetoelectric
properties}}\\
\end{center}

 \vspace{10mm}

\noindent {\sf TOM G. MACKAY}\\
School of Mathematics\\
James Clerk Maxwell Building\\
University of Edinburgh\\
Edinburgh EH9 3JZ\\
UK\\

\noindent {\sf AKHLESH LAKHTAKIA} \\
Nanoengineered Materials  Group \\
Department of Engineering Science \& Mechanics\\
212 Earth \& Engineering Sciences Building\\
Pennsylvania State University\\ University Park, PA 16802--6812\\
USA\\


\begin{abstract} The dyadic Green functions for a uniaxial
dielectric--magnetic medium, together with a reversible  field
transformation, were implemented to derive a formulation of the
Huygens principle appropriate to a uniaxial dielectric--magnetic
medium with gyrotropic--like
 magnetoelectric
properties.

\end{abstract}

\section{Introduction}

Mathematically, the Huygens principle is represented by a formula in
which the frequency--domain electromagnetic fields in a closed, source--free region are
expressed in terms of
  the tangential components of the electromagnetic fields on the
surface bounding the region. A convenient means of
expressing this formula is provided by the dyadic Green functions
appropriate to the medium filling the region  (Felsen \& Marcuvitz, 1994). Accordingly,
the Huygens principle  plays a central role in electromagnetic
scattering theory. It may be applied, for example, to the analysis
of diffraction from an aperture, wherein the aperture is formally
represented as an equivalent source (Chen, 1983); likewise, it can
be used to formulate the Ewald--Oseen extinction theorem and the T--matrix method (Lakhtakia, 1994).

While the Huygens principle for isotropic dielectric--magnetic
mediums is a staple of standard textbooks, exact formulations for
more complex mediums are generally unavailable. This general unavailability mirrors the
scarcity of exact formulations of dyadic Green functions for complex
mediums. Notable exceptions to this include the Huygens principle
for isotropic chiral mediums (Lakhtakia, 1992) and uniaxial
dielectric mediums (Bergstein \& Zachos, 1966; Lakhtakia, Varadan,
\& Varadan, 1989). A formulation of the Huygens principle for a
generally anisotropic dielectric medium was established by Ogg
(1971), but it does not yield explicit expressions as it employs a
spectral representation of the dyadic Green functions. Tai (1994)
developed a formulation of the Huygens principle for a
simply--moving isotropic dielectric medium, based on the Minkowski
constitutive relations. However, these constitutive relations
describe only nondispersive mediums and are less general than the
ones  implemented in this communication. We present here a formulation of the
Huygens principle for a  general uniaxial dielectric--magnetic
medium (Weiglhofer, 1990) with gyrotropic magnetoelectric properties (Lakhtakia \& Weiglhofer, 1990).
A harmonic time--dependence of $\exp \le - i \omega t \ri$ is implicit in the following sections.

\section{Preliminaries}

Let us consider a homogeneous  bianisotropic medium characterized by the
frequency--domain constitutive relations
\begin{equation} \l{crs}
\left.
\begin{array}{l}
\#D(\#r) = \=\eps \. \#E (\#r) + \#\Gamma \times \#H (\#r)\\
\#B(\#r) = -\#\Gamma \times \#E (\#r) + \=\mu \. \#H (\#r)
\end{array}
\right\},
\end{equation}
where the permittivity and permeability dyadics are
\begin{equation}
\left.
\begin{array}{l}
\=\eta = \epsilon \le \,  \=I - \hat{\#u} \, \hat{\#u} \, \ri + \epsilon_u
\, \hat{\#u} \, \hat{\#u}
\\
\=\mu = \mu \le \,  \=I - \hat{\#u} \, \hat{\#u} \, \ri + \mu_u
\, \hat{\#u} \, \hat{\#u}
\end{array}
\right\},
\end{equation}
with
$\=I$ being the 3$\times$3 identity dyadic\footnote{Here and
hereafter, vectors are underlined whereas dyadics are double
underlined.}. Hence, the medium exhibits uniaxial dielectric and
uniaxial magnetic properties, with the unit vector $\hat{\#u}$
indicating the direction of the distinguished axis, and the
gyrotropic--like magnetoelectric properties mediated by the vector
$\#\Gamma$. The form of the constitutive relations \r{crs} is
sufficiently general that a wide range of natural and artificial
uniaxial mediums are described by them (Mackay \& Lakhtakia, 2008).
The incorporation of gyrotropic--like magnetoelectric properties
extends the  range of mediums described by eqs. \r{crs} to include
exotic metamaterials as well as certain simply moving mediums
(Lakhtakia \& Mackay, 2006a) and  mediums inspired by
Tamm's noncovariant formulation for vacuum in
general
relativity (Lakhtakia \& Mackay, 2006b).

The frequency--domain Maxwell curl postulates can be set down as
\begin{equation} \l{mp}
\left. \begin{array}{l}
  \nabla \times \#E (\#r) + i \omega
\#\Gamma \times \#E (\#r) - i \omega \=\mu \. \#H (\#r) =  -
\#J_{\,m} (\#r)\\
 \nabla \times \#H (\#r) + i \omega \#\Gamma \times
\#H (\#r) + i \omega \=\eps \. \#E (\#r) =  \#J_{\,e} (\#r)
\end{array}
\right\},
\end{equation}
wherein $\#J_{\,e} $ and $\#J_{\,m}$ are arbitrary electric and
magnetic source current density phasors, respectively. In order to
simplify the analysis, we use the transformations (Lakhtakia \&
Weiglhofer, 1997)
\begin{equation} \l{trans}
\left.
\begin{array}{l}
\#e (\#r) = \#E (\#r) \, \exp \le i \omega \#\Gamma \. \#r \ri \\
\#h (\#r) = \#H (\#r) \, \exp \le i \omega \#\Gamma \. \#r \ri \\
\#j_{\,e} (\#r) = \#J_{\,e} (\#r) \, \exp \le i \omega \#\Gamma \. \#r \ri \\
\#j_{\,m} (\#r) = \#J_{\,m} (\#r) \, \exp \le i \omega \#\Gamma \.
\#r \ri
\end{array}
\right\}
\end{equation}
of the field and source phasors. Thereby, eqs. \r{mp} transform to
\begin{equation}
\left. \l{trans_mp}
\begin{array}{l}
  \nabla \times \#e (\#r)  - i \omega \=\mu \. \#h (\#r) =  -
\#j_{\,m} (\#r)\\
 \nabla \times \#h (\#r) + i \omega \=\eps \. \#e (\#r) =  \#j_{\,e} (\#r)
\end{array}
\right\},
\end{equation}
which yield the Helmholtzian equations
\begin{equation} \l{Helm}
\left.
\begin{array}{l}
  \les \le   \nabla \times \=I \,\ri \. \=\mu^{-1} \. \le
\nabla \times \=I \,\ri - \omega^2 \=\eps \,\ris\. \#e (\#r) =i
\omega j_{\,e} (\#r)  - \le \nabla \times \=I \,\ri \. \=\mu^{-1} \.
\#j_{\,m} (\#r) \vspace{4pt} \\
 \les \le   \nabla \times \=I \,\ri \. \=\eps^{-1} \. \le  \nabla
\times \=I \,\ri - \omega^2 \=\mu \,\ris\. \#h (\#r) =i \omega
j_{\,m} (\#r)  + \le \nabla \times \=I \,\ri \. \=\eps^{-1} \.
\#j_{\,e} (\#r)
\end{array}
\right\}.
\end{equation}

\section{Dyadic Green functions}

Let us now reproduce the dyadic Green functions for a uniaxial
dielectric--magnetic medium, as derived by
Weiglhofer (1990), for convenience as they
are essential for the analysis in the next section.

The linearity of eqs. \r{Helm}  ensures that their solution may be
expressed  as
\begin{equation}
\left.
\begin{array}{l}
 \#e (\#r) = \displaystyle{ \int \int \int \les \, \=g_{\,ee} (\#r, \#r')  \.
\#j_{\,e} (\#r') + \=g_{\,em} (\#r, \#r')  \. \#j_{\,m} (\#r') \,
\ris \; d^3 \#r'} \vspace{6pt}
\\
 \#h (\#r) = \displaystyle{ \int \int \int \les \, \=g_{\,me} (\#r, \#r')  \.
\#j_{\,e} (\#r') + \=g_{\,mm} (\#r, \#r')  \. \#j_{\,m} (\#r') \,
\ris \; d^3 \#r' } \end{array} \right\},
\end{equation}
where the integrations range over the region where $\#j_{\,e,m}$ are
not null--valued.  Herein, the 3$\times$3 dyadic Green functions $\=g_{\,
\alpha
 \beta}$ $(\alpha
 \beta =
ee, em, me, mm )$ satisfy the equations
\begin{equation} \l{gs}
\left. \begin{array}{l} i \omega \=\eps \. \=g_{\,ee} (\#r, \#r') +
\nabla \times \=g_{\,me} (\#r, \#r') = \=I \, \delta (\#r - \#r')
 \vspace{4pt} \\
\nabla \times \=g_{\,ee} (\#r, \#r') - i \omega \=\mu \. \=g_{\,me}
(\#r, \#r') = \=0 \vspace{4pt} \\
\nabla \times \=g_{\,mm} (\#r, \#r') + i \omega \=\eps \. \=g_{\,em}
(\#r, \#r') = \=0 \vspace{4pt} \\
- i \omega \=\mu \. \=g_{\,mm} (\#r, \#r') + \nabla \times
\=g_{\,em} (\#r, \#r') = - \=I \, \delta (\#r - \#r')
\end{array}
\right\},
\end{equation}
which decouple to provide the system of second order differential
equations
\begin{equation} \l{g_defs}
\left. \begin{array}{l}  \les \le   \nabla \times \=I \,\ri \.
\=\mu^{-1} \. \le \nabla \times \=I \,\ri - \omega^2 \=\eps
\,\ris\.\=g_{\,ee} (\#r, \#r') =
i \omega \, \=I \, \delta( \#r - \#r' ) \vspace{4pt} \\
  \les \le   \nabla \times \=I \,\ri \. \=\mu^{-1} \. \le \nabla
\times \=I \,\ri - \omega^2 \=\eps \,\ris\.\=g_{\,em} (\#r, \#r') =
- \le \nabla \times \=I \, \ri \.
\=\mu^{-1} \, \delta( \#r - \#r' ) \vspace{4pt} \\
 \les \le   \nabla \times \=I \,\ri \. \=\eps^{-1} \. \le  \nabla
\times \=I \,\ri - \omega^2 \=\mu \,\ris\.\=g_{\,me} (\#r, \#r') =
\le \nabla \times \=I \, \ri \.
\=\eps^{-1} \, \delta( \#r - \#r' ) \vspace{4pt} \\
 \les \le   \nabla \times \=I \,\ri \. \=\eps^{-1} \. \le  \nabla
\times \=I \,\ri - \omega^2 \=\mu \,\ris\.\=g_{\,mm} (\#r, \#r') = i
\omega \, \=I \, \delta( \#r - \#r' )
\end{array}
\right\},
\end{equation}
with $\delta$ representing the Dirac delta function. These are the
dyadic Green functions of
 a uniaxial dielectric--magnetic medium which are known to be
 (Weiglhofer, 1990)
\begin{equation} \l{G_uniaxial1} \left. \begin{array}{l}
  \=g_{\,ee} (\#r,   \#r') =
 i \omega \, \mu \les \,
   - \=T(\#r - \#r')
    +  \le \eps_u  \, \=\eps^{-1}
+ \displaystyle{\frac{ \nabla \, \nabla}{\omega^2 \eps \, \mu }} \ri
g_{\eps} (\#r -   \#r')
\, \ris \vspace{4pt} \\
  \=g_{\,em}(\#r,   \#r') =
- \eps   \, \=\eps^{-1}  \.\le \nabla \times \=I \,\ri  \.\les \mu_u
\, g_{\mu} (\#r - \#r') \, \=\mu^{-1}  + \=T(\#r - \#r')\, \ris
\vspace{4pt}
  \\
  \=g_{\,me}(\#r,   \#r') =
- \mu  \, \=\mu^{-1} \.\le \nabla \times \=I \,\ri \.\les - \eps_u
\, g_{\eps} (\#r - \#r') \, \=\eps^{-1}  + \=T(\#r - \#r')\, \ris
\vspace{4pt}
\\
  \=g_{\,mm} (\#r,   \#r') =
 i \omega \, \eps \les \,  \=T(\#r - \#r')
    +  \le \mu_u  \, \=\mu^{-1}
+ \displaystyle{\frac{ \nabla \, \nabla}{\omega^2 \eps  \, \mu  }}
\ri g_{\mu} (\#r - \#r') \, \ris
\end{array} \right\}.
\end{equation}
The scalar Green functions $g_{\eps, \mu}$ in eqs. \r{G_uniaxial1}
are defined by
\begin{equation}
g_{\eta}  (\#R) = \frac{\exp \les i \omega \le \eps \, \mu \ri^{1/2}
\, \le \, \eta_{u}  \, \#R\. \=\eta^{-1} \.\#R \, \ri^{1/2} \,
\ris}{4 \pi
  \le \, \eta_u   \,
\#R\. \=\eta^{-1} \.\#R \, \ri^{1/2}}, \qquad \quad (\eta = \eps,
\mu ), \l{ge_gm}
\end{equation}
while the dyadic function $\underline{\underline{T}}$ is specified
as
\begin{eqnarray}
\=T(\#R) &=& \nonumber \frac{\le \, \#R \times \hat{\#u} \, \ri
 \le \, \#R \times \hat{\#u} \, \ri}{  \le \, \#R \times \hat{\#u} \,
 \ri^2} \les \,  \frac{\eps_u }{\eps } g_{\eps} (\#R) - \frac{\mu_{u} }{\mu
} g_{\mu} (\#R) \, \ris
\\ && + \frac{g_{\eps} (\#R) \, \le \eps_{u}  \, \#R\.
\=\eps^{-1}  \.\#R \, \ri^{1/2}  - g_{\mu} (\#R)
 \,  \le \mu_{u}   \, \#R\. \=\mu^{-1}  \.\#R \, \ri^{1/2}}{i \omega \, \le \, \eps  \, \mu
\, \ri^{1/2} \le \, \#R \times \hat{\#u} \, \ri^2}\,
 \nonumber \\ && \times \les \, \=I - \hat{\#u}
\, \hat{\#u} -
  \frac{
2 \le \, \#R \times  \hat{\#u} \, \ri \le \, \, \#R \times \hat{\#u}
\, \ri} { \le \, \#R \times  \hat{\#u} \, \ri^2}\,\ris .
  \l{dyadic_T}
\end{eqnarray}
On considering the dyadic transposes, which we denote by the
superscript `T', eqs. \r{G_uniaxial1} reveal the symmetries
\begin{equation} \l{sym1}
\left.
\begin{array}{l}
\=g^T_{\,ee} (\#r, \#r') = \=g_{\,ee} ( \#r', \#r) \vspace{4pt} \\
\=g^T_{\,em} (\#r, \#r') = - \=g_{\,me} ( \#r', \#r) \vspace{4pt} \\
\=g^T_{\,me} (\#r, \#r') = - \=g_{\,em} ( \#r', \#r) \vspace{4pt} \\
\=g^T_{\,mm} (\#r, \#r') = \=g_{\,mm} ( \#r', \#r)
\end{array}
\right\},
\end{equation}
 while  interchanging
$\#r$ and $\#r'$ yields
\begin{equation} \l{sym2}
\left.
\begin{array}{l}
\=g_{\,ee} (\#r, \#r') = \=g_{\,ee} (\#r', \#r) \vspace{4pt} \\
\=g_{\,em} (\#r, \#r') = - \=g_{\,em} (\#r', \#r) \vspace{4pt} \\
\=g_{\,me} (\#r, \#r') = - \=g_{\,me} (\#r', \#r) \vspace{4pt} \\
\=g_{\,mm} (\#r, \#r') =  \=g_{\,mm} (\#r', \#r)
\end{array}
\right\}.
\end{equation}

\section{Huygens principle}

\begin{figure}[!ht]
\centering \psfull \epsfig{file=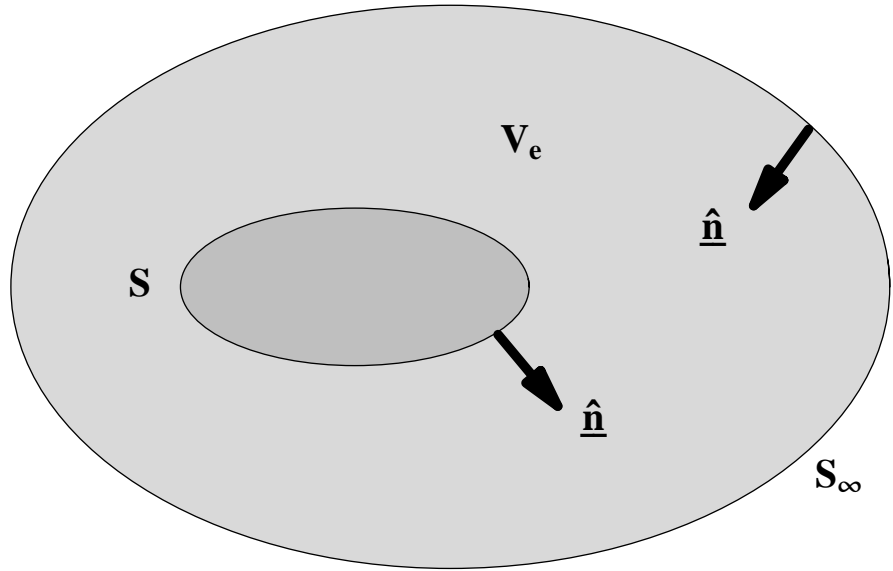,width=5.0in}
  \caption{\label{fig1} The bounded region $V_e$ enclosed by the surfaces $S$ and $S_\infty$. The unit vector
   $\hat{\#n} $ on $S \cup S_\infty$ is directed  into
$V_e$.}
\end{figure}

The dyadic Green functions \r{G_uniaxial1} may be exploited to
derive mathematical statements of the Huygens principle pertaining
to radiation and scattering in a uniaxial dielectric--magnetic
medium with gyrotropic--like
 magnetoelectric
properties. Consider a source--free region $V_e$, which is occupied
by a medium described by the constitutive relations \r{crs} and
enclosed by the finite  surfaces $S$ and $S_\infty$, as
schematically illustrated in  Fig.~\ref{fig1}.  The source phasors
$\#J_{\,e,m}$ (and $\#j_{\,e,m}$ ) are nonzero only inside $S$. The
unit vector $\hat{\#n}$ on $S \cup S_\infty$ is directed into $V_e$.

We introduce the dyadic function
\begin{equation}
\=A (\#r, \#r' )= - \#e (\#r) \times \lec \, \=\mu^{-1} \. \les
\nabla \times \=g_{\,ee} (\#r, \#r' ) \ris \ric - \lec \=\mu^{-1} \.
\les \nabla \times \#e (\#r) \ris \ric \times \=g_{\,ee} (\#r, \#r'
)\,;
\end{equation}
and note that
\begin{equation}
\nabla \. \=A (\#r, \#r' ) = \#e (\#r) \. \le \nabla \times \lec
\=\mu^{-1} \. \les \nabla \times \=g_{\,ee} (\#r, \#r') \ris \ric
\ri - \le \nabla \times \lec \=\mu^{-1} \. \les \nabla \times \#e
(\#r) \ris \ric \ri \. \=g_{\,ee} (\#r, \#r'),
\end{equation}
by virtue of the identity
\begin{equation}
\nabla \. \le \, \#p \times \=Q \, \ri= \le \, \nabla \times \#p
\,\ri \. \=Q - \#p \. \le \, \nabla \times \=Q \, \ri,
\end{equation}
and the fact that the dyadic $\=\mu^{-1}$ is symmetric. The dyadic
divergence theorem (Tai, 1994; van Bladel, 1985)
\begin{eqnarray}
&& \int \int \int_{V_e}  \nabla \. \=A (\#r, \#r' ) \; d^3 \#r = -
\int \int_{S \cup S_\infty} \, \hat{\#n} (\#r) \. \=A (\#r, \#r' )
\; d^2 \#r
\end{eqnarray}
then yields
\begin{eqnarray} \l{e1}
 \int \int \int_{V_e} \les  \#e (\#r) \. \le \nabla \times \lec
\=\mu^{-1} \. \les \nabla \times \=g_{\,ee} (\#r, \#r') \ris \ric
\ri - \le \nabla \times \lec \=\mu^{-1} \. \les \nabla \times \#e
(\#r) \ris \ric \ri \. \=g_{\,ee} (\#r, \#r') \ris \; d^3 \#r && \nonumber \\
=  \int \int_{S \cup S_\infty} \Big[ \les \hat{\#n} (\#r) \times \#e
(\#r) \ris \. \lec \, \=\mu^{-1} \. \les \nabla \times \=g_{\,ee}
(\#r, \#r' ) \ris \ric \hspace{30mm} && \nonumber \\  + \le
\hat{\#n} (\#r) \times \lec \=\mu^{-1} \. \les \nabla \times \#e
(\#r) \ris \ric \ri \. \=g_{\,ee} (\#r, \#r' ) \Big]\; d^2
\#r,\hspace{10mm} &&
\end{eqnarray}
after using the identity $\#s \. \le \#p \times \=Q \ri = \le \#s \times
\#p \ri \. \=Q$. Now  eq. \r{Helm}$_1$ and
\r{g_defs}$_1$ can be employed, together with the
symmetry of the constitutive dyadic $\=\eps$ and the fact that there
are no sources in $V_e$, to evaluate the integral on the left side
of eq. \r{e1}. Thus, we find
\begin{eqnarray} \l{e2}
i \omega \,  \#e (\#r') & = & \int \int_{S \cup S_\infty} \Big[ \les
\hat{\#n} (\#r) \times \#e (\#r) \ris \. \lec \, \=\mu^{-1} \. \les
\nabla \times \=g_{\,ee} (\#r, \#r' ) \ris \ric \nonumber \\ && +
\le \hat{\#n} (\#r) \times \lec \=\mu^{-1} \. \les \nabla \times \#e
(\#r) \ris \ric \ri \. \=g_{\,ee} (\#r, \#r' ) \Big]\; d^2 \#r,
\qquad \#r' \in V_e.
\end{eqnarray}
Next, let $S_\infty$ be sufficiently  distant from $S$
that the
 integral herein on the surface $S_\infty$ is eliminated by virtue
 of  satisfaction of appropriate radiation conditions by the
 dyadic Green functions (Felsen \& Marcuvitz, 1994).
Equations \r{trans_mp}$_1$ and    \r{gs}$_2$
allow the  curl terms in the integrand of eq. \r{e2} to be replaced
by explicit terms; hence,
\begin{equation} \l{e3}
  \#e (\#r')  =   \int \int_{S} \lec \les \hat{\#n}
(\#r) \times \#e (\#r) \ris  \. \=g_{\,me} (\#r, \#r' ) + \les
\hat{\#n} (\#r) \times \#h (\#r) \ris  \. \=g_{\,ee} (\#r, \#r' )
\ric\; d^2 \#r, \qquad \#r' \in V_e.
\end{equation}
On interchanging $\#r$ and $\#r'$, and exploiting the symmetries of
$\=g_{\,ee}$, $\=g_{\,em}$ and $\=g_{\,me}$ listed in eqs. \r{sym1}
and \r{sym2}, it emerges that
\begin{equation} \l{hp0}
  \#e (\#r)  =   \int \int_{S} \lec
\,   \=g_{\,ee} (\#r, \#r' ) \. \les \hat{\#n} (\#r') \times \#h
(\#r') \ris  -
  \=g_{\,em} (\#r, \#r' ) \.
  \les \hat{\#n}
(\#r') \times \#e (\#r') \ris  \ric\; d^2 \#r', \qquad \#r \in V_e.
\end{equation}
And, lastly, after inverting the transformation of $\#E$ presented
in eq. \r{trans}$_1$, the Huygens principle for the electric field
phasor emerges as
\begin{eqnarray} \l{hp1}
  \#E (\#r) & = &    \int \int_{S} \lec
\,   \=g_{\,ee} (\#r, \#r' ) \. \les \hat{\#n} (\#r') \times \#H
(\#r') \ris  -
  \=g_{\,em} (\#r, \#r' ) \.
  \les \hat{\#n}
(\#r') \times \#E (\#r') \ris  \ric \nonumber \\ && \times \exp \les
- i \omega \#\Gamma \. \le \#r - \#r' \ri \ris  \; d^2 \#r',
\hspace{60mm} \#r \in V_e.
\end{eqnarray}

To find the corresponding expression for the magnetic field phasor, we take
the curl of both sides of eq. \r{hp0} and utilize eqs. \r{gs}$_{2,3}$
to get
\begin{eqnarray} \l{hp2}
\nabla \times  \#e (\#r)  &=&   \int \int_{S} \Big( \, i \omega
\=\mu \. \=g_{\,me} (\#r, \#r' ) \. \les \hat{\#n} (\#r') \times \#h
(\#r') \ris  \nonumber \\ && + \frac{1}{i \omega}  \nabla \times
\lec \=\eps^{-1} \. \les \nabla \times \=g_{\,mm} (\#r, \#r') \ris
\ric
   \.
  \les \hat{\#n}
(\#r') \times \#e (\#r') \ris  \Big)\; d^2 \#r', \qquad \#r \in V_e.
\end{eqnarray}
Then, exploitation of eqs. \r{trans_mp}$_1$ and
 \r{g_defs}$_4$  to eliminate the
curl terms delivers
\begin{eqnarray} \l{hp3}
 \#h (\#r)  &=&   \int \int_{S} \lec  \,
\=g_{\,me} (\#r, \#r' ) \. \les \hat{\#n} (\#r') \times \#h (\#r')
\ris  -   \=g_{\,mm} (\#r, \#r')
   \.
  \les \hat{\#n}
(\#r') \times \#e (\#r') \ris  \ric\; d^2 \#r', \nonumber \\ &&
\hspace{90mm}  \#r \in V_e, \; \#r \notin S,
\end{eqnarray}
wherein the observation point $\#r$ is explicitly excluded from the
surface of the source region.  The Huygens principle for the
magnetic field phasor thus emerges as
\begin{eqnarray} \l{hp4}
 \#H (\#r)  &=&  
\int \int_{S} \lec  \, \=g_{\,me} (\#r, \#r' ) \. \les \hat{\#n}
(\#r') \times \#H (\#r') \ris  -   \=g_{\,mm} (\#r, \#r')
   \.
  \les \hat{\#n}
(\#r') \times \#E (\#r') \ris  \ric \nonumber \\ && \times \exp \les
- i \omega \#\Gamma \. \le \#r - \#r' \ri \ris \; d^2 \#r',
\hspace{50mm}  \#r \in V_e, \; \#r \notin S,
\end{eqnarray}
upon
 inverting the transformation of $\#H$ presented in eq.
\r{trans}$_2$.

\section{Closing remarks}

The main results of this communication are  eqs. \r{hp1} and
\r{hp4}, which represent formulations of the Huygens principle
appropriate to the bianisotropic medium described by the
constitutive relations \r{crs}. This is the most general formulation
of the Huygens principle presently available, as far as we are
aware. We note that the formulation of the Huygens principle for a
uniaxial dielectric medium derived by Lakhtakia, Varadan, \& Varadan
(1989) follows immediately from eqs. \r{hp1} and \r{hp4} upon
setting $\#\Gamma = \#0$ and $\=\mu = \muo \=I$ , where $\muo$ is
the permeability of free space.

\vspace{10mm}

 \noindent{\bf References}\\

\noindent Bergstein, L., and T. Zachos.  1966. A Huygens' principle
for uniaxially anisotropic media. \emph{J. Opt. Soc. Am.}
56:931--937.\\

\noindent Chen, H.C. 1983. \emph{Theory of electromagnetic waves}.
New York, NY, USA:  McGraw--Hill.\\

\noindent Felsen, L.B.,  and N. Marcuvitz. 1994. \emph{Radiation and
scattering of waves}. Piscataway, NJ, USA: IEEE Press.\\

\noindent Lakhtakia, A. 1992.
On the Huygens's principles and the Ewald--Oseen extinction theorems for, and the scattering of, Beltrami fields.
\emph{Optik} 91:35--40.\\

\noindent Lakhtakia, A. 1994. \emph{Beltrami fields in chiral
media}. Singapore: World Scientific.\\

\noindent Lakhtakia, A., and T.G. Mackay. 2006a. Simple derivation
of dyadic Green functions of a simply moving, isotropic
dielectric--magnetic medium. \emph{Microw. Opt. Technol. Lett.}
48:1073--1074.\\

\noindent Lakhtakia, A., and T.G. Mackay. 2006b. Dyadic Green
function for an electromagnetic medium inspired by general
relativity. \emph{Chin. Phys. Lett.} 23:832--833. \\

\noindent Lakhtakia, A., V.K. Varadan, and V.V. Varadan. 1989. A
note on Huygens's principle for uniaxial dielectric media. \emph{J.
Wave--Mater. Interact.} 4:339--343.\\

\noindent Lakhtakia, A., and W.S. Weiglhofer. 1997. On
electromagnetic fields in a linear medium with gyrotropic--like
magnetoelectric properties. \emph{Microw. Opt. Technol. Lett.}
15:168--170. \\

\noindent Mackay, T.G., and  A. Lakhtakia. 2008.
 Electromagnetic fields in linear bianisotropic mediums.
 \emph{Prog. Opt.} 51:121--209. \\

\noindent Ogg, N.R. 1971. A Huygen's principle for anisotropic
media. \emph{J. Phys. A: Gen. Phys.}  4:382--388. \\

\noindent Tai,  C.T. 1994. \emph{Dyadic Green functions in
electromagnetic theory, 2nd ed}. Piscataway, NJ, USA: IEEE Press; p. 272. \\

\noindent van Bladel, J. 1985. \emph{Electromagnetic fields}.
Washington, DC, USA: Hemisphere; p. 509. \\

\noindent Weiglhofer, W.S. 1990. Dyadic Green's functions for
general uniaxial media. \emph{IEE Proc., Pt. H} 137:5--10.\\

\newpage

\end{document}